\documentclass[twocolumn]{aastex61}
\usepackage{amsmath,amstext}
\usepackage[T1]{fontenc}
\usepackage{apjfonts}
\usepackage[figure,figure*]{hypcap}
\usepackage{graphicx}

\shorttitle{VdBH 222}
\shortauthors{R.\, Asa'd, et al.}

\begin{document}

\title{Analysis of Red-Supergiants in VdBH 222}


\author{Randa Asa'd}
\affiliation{American University of Sharjah, Physics Department, P.O. Box 26666, Sharjah, UAE}

\author{M. Kovalev}
\affiliation{Max Planck Institute for Astronomy, Königstuhl 17, 69117 Heidelberg, Germany}

\author{B. Davies}
\affiliation{Astrophysics Research Institute, Liverpool John Moores University, Liverpool Science Park ic2, 146 Brownlow Hill, Liverpool L3 5RF, UK}

\author{V. D. Ivanov}
\affiliation{European Southern Observatory, Karl-Schwarzschild-Straße 2, D-85748 Garching bei M\"{u}nchen, Germany}

\author{M. Rejkuba}
\affiliation{European Southern Observatory, Karl-Schwarzschild-Straße 2, D-85748 Garching bei M\"{u}nchen, Germany}

\author{A. Gonneau}
\affiliation{Institute of Astronomy, University of Cambridge, Madingley Road, Cambridge CB3 0HA, United Kingdom}

\author{S. Hernandez}
\affiliation{Space Telescope Science Institute, 3700 San Martin Drive, Baltimore, MD 21218, USA}

\author{C. Lardo}
\affiliation{Laboratoire d'astrophysique, \'{E}cole Polytechnique F\'{e}d\'{e}rale de Lausanne (EPFL), Observatoire de Sauverny, CH-1290, Versoix, Switzerland}

\author{M. Bergemann}
\affiliation{Max Planck Institute for Astronomy, Königstuhl 17, 69117 Heidelberg, Germany}

\correspondingauthor{Randa Asa'd}
\email{raasad@aus.edu}

\begin{abstract}

 Recent surveys uncovered new Young Massive Clusters (YMCs) that host dozens of Red Supergiants (RSGs) in the inner Milky Way.
 These clusters are ideal for studying the most recent and violent star formation events in the inner Galaxy. However, due to the high extinction that affects the Galactic plane, they need to be studied through infrared (IR) spectroscopy. IR spectra of RSGs have proven to be powerful tools for obtaining chemical abundances. \\
 We present the first [Fe$/$H] measurement ($-$0.07$\pm$0.02) for the YMC VdBH 222 through analysis of its RSGs using VLT/X-shooter spectra. We find no evidence for Multiple Stellar Populations (MSPs) in this YMC, contrary to what is routinely observed in older massive clusters.

\end{abstract}


\section{Introduction and Motivation}

Accurate information about the abundances of different elements reveal information about the star formation and chemical enrichment history of the host galaxy \citep[i.e.][and references therein]{Matteucci12}.
Additionally, determining the abundances of elements resulting from different nucleosynthesis processes such as Type I supernovae (producing Fe-peak elements: Sc, V, Cr, Mn, Fe, Co and Ni), core collapse supernovae (producing $\alpha$-elements: O, Ne, Mg, Si S, Ar, Ca and Ti) and winds from evolved stars can provide us with information about these complex mechanisms.\\
The chemical enrichment history of the Galactic center is
a crucial matter to unveil the Galaxy's evolution.

The Galactic center has experienced a recent rich star-forming activity which formed numerous young star clusters dominated by red supergiants and large number of massive stars observed there. The origin of this activity can be traced by the metallicity and chemical abundances of these young objects \citep{Figer99, Figer02, Stolte14}. \\
Star clusters lie at the heart of modern astrophysics because they can provide information both about the chemical evolution of their host galaxy as well as the stellar evolution of their constituent stars \citep{Bertelli03, Brodie06, Gratton19}.
Studies of massive star clusters of different ages has received further boost after the discovery of ubiquitous presence of Multiple Stellar Populations (MSP) phenomenon, defined as star-to-star variation of light element (He, C, N, O, Na, Al) abundances, in old Milky Way clusters \citep[e.g.][]{Carretta09}, which has changed our view of star clusters as simple stellar populations. The origin of the MSP phenomenon is still poorly understood. For the recent review see \citet{Bastian18}.\\
\citet{Larsen06} analyzed the chemical abundances and abundance patterns with near-infrared spectroscopy in a young cluster.
Several studies \citep{Davies10, Davies15, Patrick16} demonstrated that the IR spectra of red supergiants (RSGs) can be used to measure abundances. \citet{Gazak14} showed that the technique works down to resolving powers of R = 3000.
RSGs are cool ($\sim$ 4000 K), highly luminous ($\sim$ $10^5$  L$_\odot$) stars of spectral class K and M with masses between 10 to 30 M$_\odot$. They are evolved stars that left the main sequence but they are young ($<$20 Myr) and have short lifetimes \citep{Levesque10}.\\
\cite{Cabrera-Ziri16, Lardo17} used the integrated spectra of Young Massive Clusters (YMC) in the J-band (1.1 to 1.4 $\mu$m) which is dominated by RSG to search for evidence for MSP in star of different masses (while previous studies mostly focused on low-mass stars).
Studies covering a broader sample of clusters will help answering long standing questions like: are YMCs and Globular Clusters (GCs) objects of the same nature? Do both share the peculiar abundance patterns?\\
In this work we analyse the YMC VdBH 222 (SIMBAD identifier: Cl VDBH 222), located in the inner Milky Way, found by \citet{vandenBergh75}. This cluster is ideal for this type of study as it is one of the few stellar clusters with confirmed RSGs. Due to the high extinction that affects the Galactic plane, stars in VdBH 222 need to be studied through IR spectroscopy. \citet{Piatti02} estimated an age of 60$\pm$30 Myr for it and concluded that it would be important to derive the cluster metallicity in order to improve our knowledge of the radial metal abundance gradient and the age-metallicity relation for the disk of our Galaxy.\\
\citet{Marco14} characterised this cluster using a comprehensive set of multi-wavelength observations and determined a reddening E(B-V) of 2.45 $\pm$ 0.15 using non-standard extinction law with a value of R = 2.9 and an age range of 12$-$16 Myr at a distance of 7$-$10 kpc. They confirmed that this YMC has a likely mass of 2x10$^4$M$_\odot$, and that it is an extremely compact cluster with very few members lying outside a radius of 1.5' from the cluster's center.
They pointed that VdBH 222 is much closer to the Galactic centre (and so to the nominal tip of the bar) than other massive clusters reported in the area \citep{Davies12, Ram14}.
\citet{Clark15} inferred a distance of about 6kpc for VdBH 222 and an age of about 20 Myr.\\
\citet{Marco14} identified nine RSG in this cluster and derived an average radial velocity v$_{LSR} = -$99$\pm$4 km s$^{-1}$. In their analysis of open cluster kinematics with Gaia DR2, \citet{2018Gaia} derived a radial velocity of $-$119.3$\pm$2.8 km s$^{-1}$ based on five stars.\\
Using X-shooter at the Very Large Telescope (VLT) IR data, we provide the first estimate of the properties and iron abundance of six RSG in this cluster. We also search for evidence of MSP through analysis of [Al/Fe].\\
The paper is organized as follows: in section 2 we describe our observations and the data reduction. In section 3 we discuss the models and the method used, in section 4 we provide details about the error analysis and in section 5 we present our results. In section 6 we shed light on the Multiple Stellar Populations phenomenon in this cluster. The conclusion is provided in section 7.

\section{Observations and Data Reduction}

We obtained the IR spectra of six RSGs in VdBH 222 using VLT/X-shooter \citep{D'Odorico06, Vernet11} in service mode under ESO programme number 0103.D-0881(A) (PI R. Asad). These observations provide continuous spectral coverage from  0.3$-$2.4 $\mu$m.
Integration times were chosen to achieve a signal-to-noise ratio (S/N) of at least 100 in the NIR. By examining the color-magnitude diagrams (CMDs) of the cluster we established that the RSGs are bright having K~6-8 mag. Hence with 2x30s integration in the NIR we were able to get S/N of a few hundreds.
 The sample of RSG stars for the cluster was assembled from \citet{Marco14}.
We used the slits 1.0", 0.9", 0.9" for UVB, VIS and NIR channels, respectively. This gives a resolution higher that R$\sim$5600 in the NIR arm. The precise value of R for each RSGs is determined at the analysis stage.
For this work we used the J-band spectra, that are dominated by atomic rather than
molecular absorption, allowing accurate stellar
abundances measurements \citep{Davies10, Davies15}.
Details about the S$/$N of each cluster RSG target is given in Table \ref{T1}. The average S$/$N of the clusters targets is 233.

We started with the advanced data products for our observed sample from the ESO archive. The spectra were processed by applying the standard spectroscopic data reduction steps\footnote{www.eso.org/observing/dfo/quality/PHOENIX/XSHOOTER/processing.html} as described in the associated data release description available on the ESO Phase 3 website\footnote{http://www.eso.org/rm/api/v1/public/releaseDescriptions/70}.\\
We corrected the 1D extracted and flux-calibrated spectra for telluric absorption using \texttt{molecfit} \citep{Smette15,Kausch15}. We used the same approach as developed for the X-shooter Spectral Library \citep{Gonneau20}. First we apply \texttt{molecfit} to the entire near-IR spectrum to derive the precipitable water vapour column (PWV). Then we divide the spectrum into smaller wavelength segments and apply \texttt{molecfit} locally using the PWV value determined before. The corrected wavelength segments are then merged together.

\section{Models and Method}

The kinematic parameters and chemical abundances were derived by comparing the observed spectra with synthetic spectra generated by
\textit{the~Payne} spectral model \citep{ting19}, through a $\chi^2$ minimization in the wavelength range $11600-12200$\,\AA.
As a training set for \textit{the~Payne} we adopted model atmospheres calculated with MARCS code \citep{Gustafsson08}. The NLTE spectral grids were computed by \citet{bergemann15}, using NLTE departures for Si, Ti and Fe lines from \citet{bergemann12, Bergemann12b, Bergemann13}. This model is similar to the model used in \citet{Davies15}\\
The parameter space includes stellar parameters (T$_{eff}$, log(g), microturbulence ($\xi$) and [Fe/H]) together with radial velocity, resolution and normalization coefficients for Chebyshev polynomials.
The grid of models is computed for a range of  T$_{eff}$ between 3400 and 4400 K in steps of 100K, log(g) between $-$1.0 and $+$1.0 in steps of 0.25 (cgs units), $\xi$ from 1 to 5 km s$^{-1}$ and [Fe/H] between $-$ 1.0 and $+$1.0 dex is steps of 0.25 dex.
The synthetic spectrum was shifted to radial velocity and degraded to the spectral resolution (using Gaussian filter) of observed stellar spectrum, which was normalized using linear combination of the first four Chebyshev polynomials \citep[similar to][]{kovalev19}.

\section{Error Analysis}

In addition to the statistical errors associated with the fitting, we discuss the following errors: RMS from tests done on simulated mock stars and degeneracy errors. The total errors are taken to be the quadrature sum of the statistical errors from $\chi^2$ minimisation and the RMS from tests on simulated mock stars.

\subsection {Tests on simulated mock stars}

\begin{figure*}
\includegraphics[width=\textwidth]{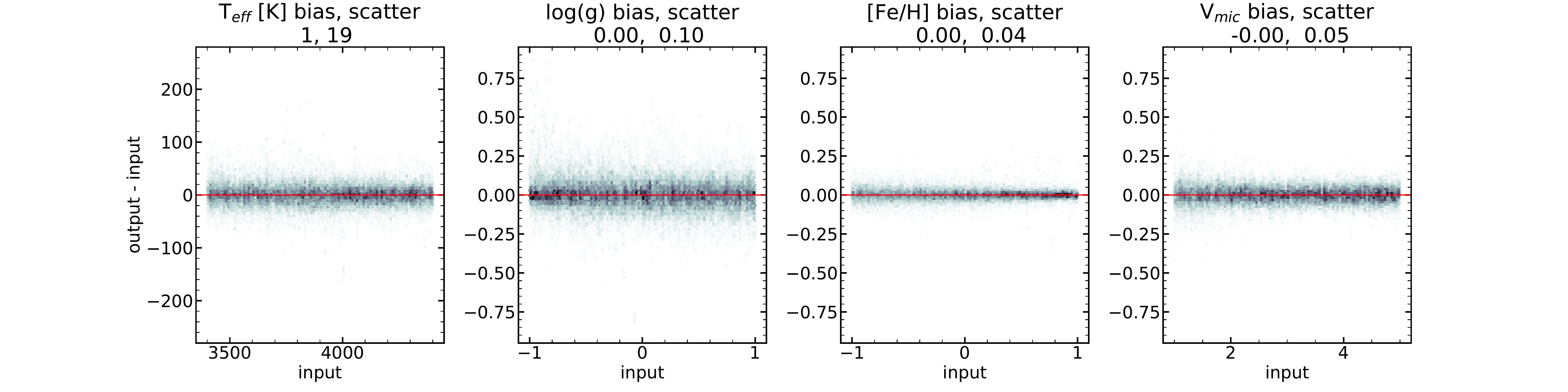}
\caption{The distribution of the derived parameters for the set of mock stars. The black dots represent the sample of 100,000 mock stars, the red line is the central line. The bias and RMS are shown in panels' titles.}
\label{Fig3}
\end{figure*}

We use \textit{the~Payne} spectral model to create 100,000 mock stellar spectra (generated in random points within parameter's space of original model grid) with Gaussian noise corresponding to $S/N=100$.  All mock spectra were degraded to a resolution of $R=9700$ and shifted to random radial velocities from $-40$ to $40$ km s$^{-1}$. They were also multiplied by a polynomial function representing continuum placement uncertainty. \\
We then aimed to recover the original stellar parameters.
The 4 panels of Figure \ref{Fig3} show the offset and scatter for each (output-input) versus (input) parameter. We use the bias (offset) from the expected value and RMS as the measure of the fitting performance. All parameters show negligible bias values. The RMS value is 19 K in $T_{eff}$, 0.1 dex in log(g), 0.04 dex in metallicity and 0.05 km s$^{-1}$ in $\xi$.

\subsection{Degeneracy}

 To investigate the uncertainty caused by the degeneracies between estimated parameters, 
we calculate the $\chi^2$ values for $40000$ points around optimal solution, changing two parameters at time and keeping the other parameters fixed to the optimal values. The contour lines are shown in Figure \ref{Fig2b} for values corresponding to 1, 2, 3 $\sigma$ levels. Optimal parameters are shown with pluses.\\
T0, T1 and T2 are the coefficients of Chebyshev polynomials used in normalisation.
The figures are in agreement with the degeneracy analysis of \citet{Davies15} who emphasized the degeneracy between [Fe/H] and log(g).
Those uncertainties are already included in the statistical errors from $\chi^2$, because we fit for all parameters simultaneously. After each fit we obtain a covariance matrix from which we calculate the quadrature sum of diagonal elements as statistical errors. \\

\section{Results}

\begin{figure*}
\plotone{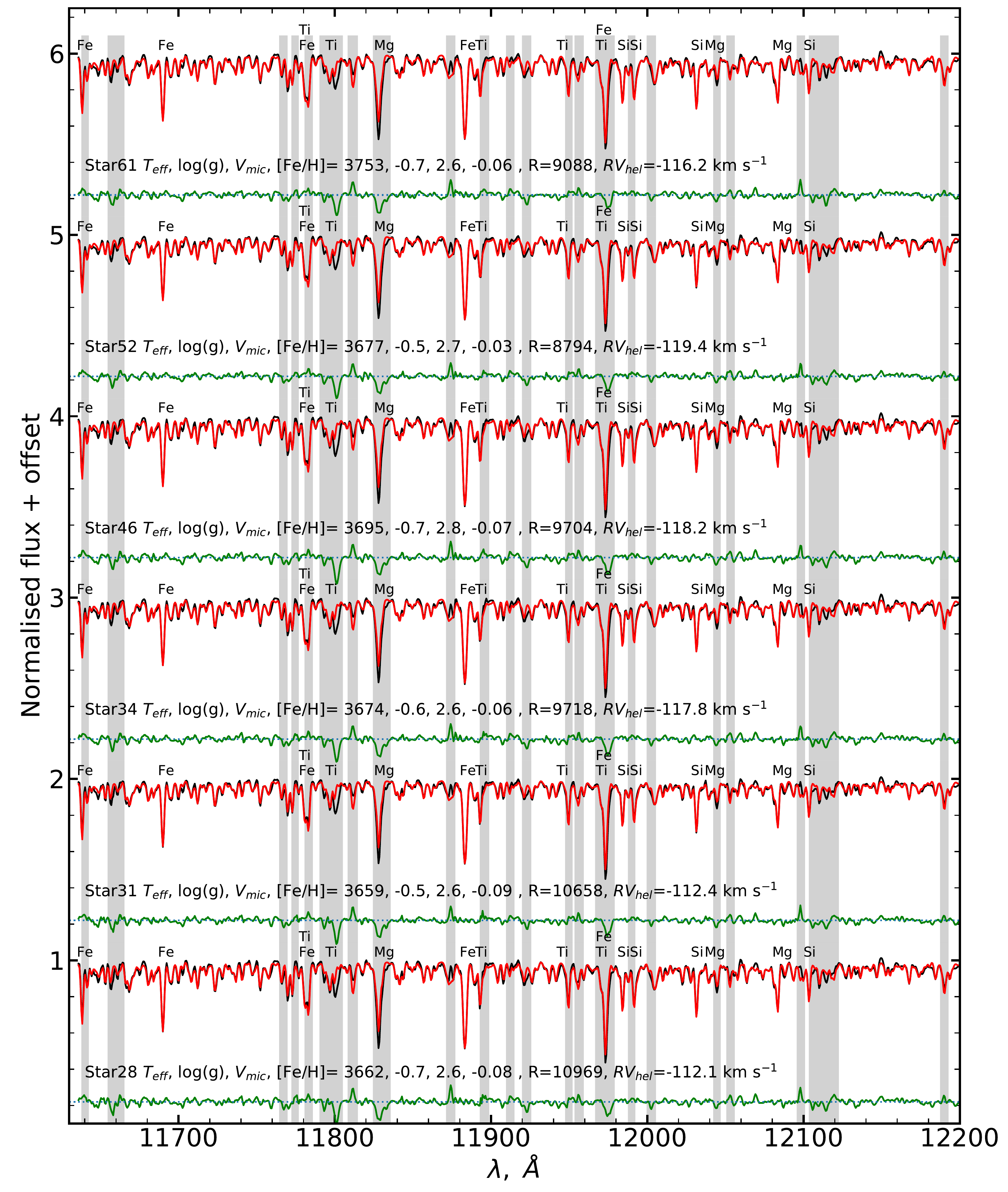}
\caption{The best match between the observed RSGs (black) and NLTE model (red) masking the regions of 1\AA $ $ around points where the mean residuals are greater than 0.03 (grey shaded). Residuals are shown in green.  }
\label{Fig1b}
\end{figure*}

Figure \ref{Fig1b} shows the best match between the observed RSGs and NLTE MARCS models when masking the regions of 1\AA $ $ around points where the mean residuals are greater than 0.03. The numerical results are listed in Table \ref{T1}. The ID of each star reflects the unique OB number given during the observing run. \footnote{All stars in our sample start with OB22669, we only used the last two digits which are unique per star. Star34 for example, is OB2266934.}

All RSGs have radial velocities within $\pm$3km s$^{-1}$ from each other which confirms that they are all members of the cluster.
The average radial velocity for our sample is $-$116.5$\pm$3 km s$^{-1}$ which is consistent with the two values from other studies, $-$119.3$\pm$2.8 km s$^{-1}$ from \citet{2018Gaia} based on their open cluster kinematics with Gaia DR2 and $-$99$\pm$4 km s$^{-1}$ obtained by \citet{Marco14}.

Star31 has the lowest temperature. Star61 has the highest temperature. The overall sequence in temperature values is consistent with the spectral classes identified for this sample by \citet{Marco14}.

\begin{figure}
\plotone{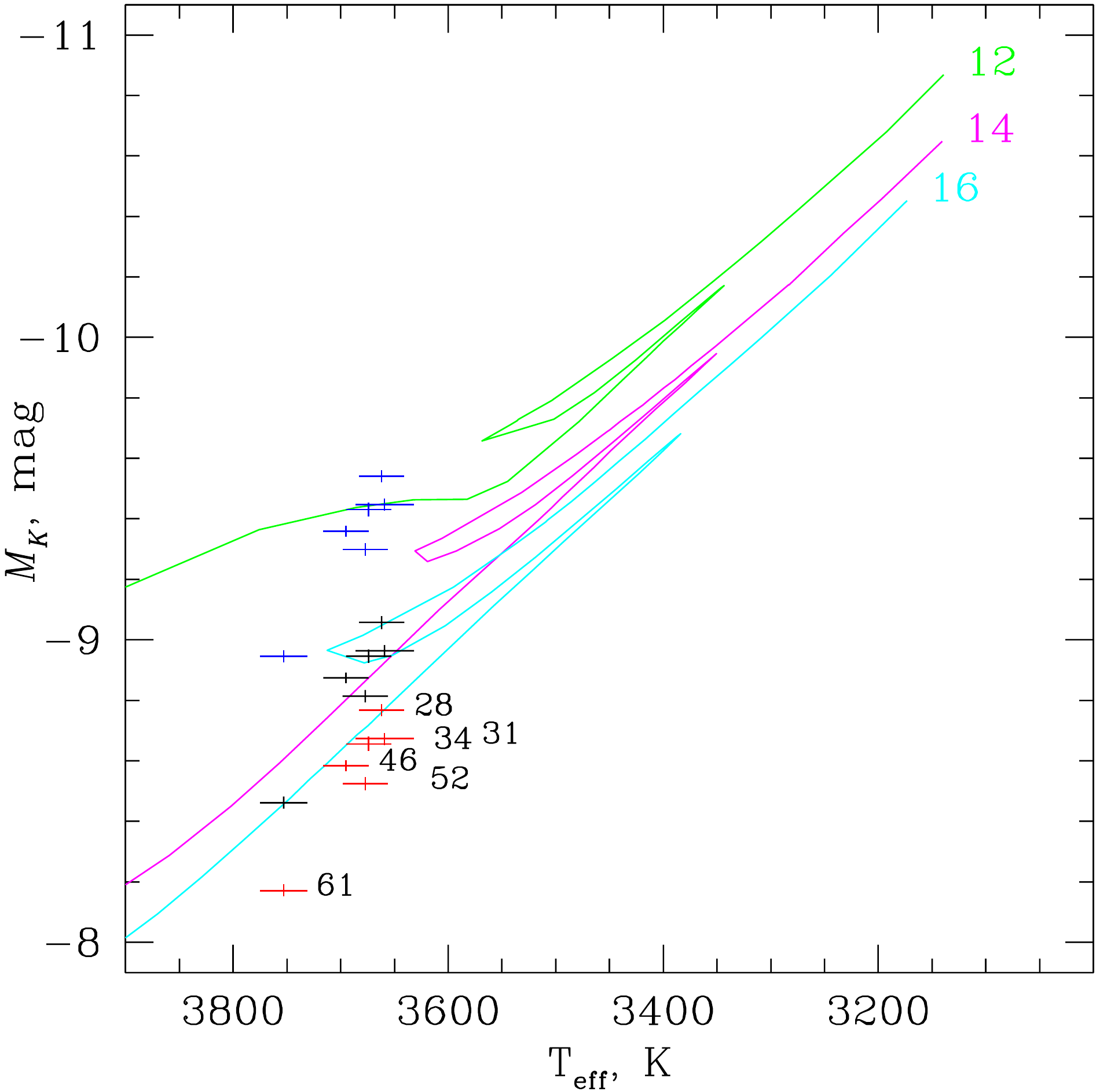}
\caption{HRD for VdBH 222 using our RSG sample. The red, black and dark blue crosses correspond to distances of 7, 8 and 10 kpc, respectively. We use 12Myr, 14Myr and 16Myr isochrones with E(B$-$V)=2.45.}
\label{cmd}
\end{figure}

In figure \ref{cmd} we plot the Hertzsprung–Russell diagram (HRD) for this cluster using the temperatures we obtained for the RSGs in our sample. The NIR magnitudes are form the 2MASS catalogue.
We use Padova\footnote{http://stev.oapd.inaf.it/cgi-bin/cmd} \citep{Bressan12, Marigo13, Pastorelli19} 12Myr, 14Myr and 16 Myr isochrones with E(B$-$V) = 2.45 \citep[values from ][]{Marco14}.
The red, black and dark blue crosses correspond to distances of 7, 8 and 10 kpc respectively. The comparison of the new stellar parameters with the theoretical models favour older ages and shorted distances or $\sim$ 16Myr and D$\sim$8 kpc.

RSGs in our sample have temperatures in the 3650$-$3750 K range, surface gravity log(g) from $-$0.7 to $-$0.5 and microturbulence in the 2.6$-$2.8 km s$^{-1}$ range.

We find the average [Fe$/$H] to be $-$0.07$\pm$0.02. This is the first estimation of metallicity in this cluster.  This value is in line with findings from other studies of RSGs in inner Milky Way \citep[i.e. see discussions in][]{Cunha07, Ramirez00, Carr00, Davies09B, Davies09, Origlia13, Origlia16, Origlia19}.



\section{Multiple Stellar Populations}

\begin{figure*}
\plotone{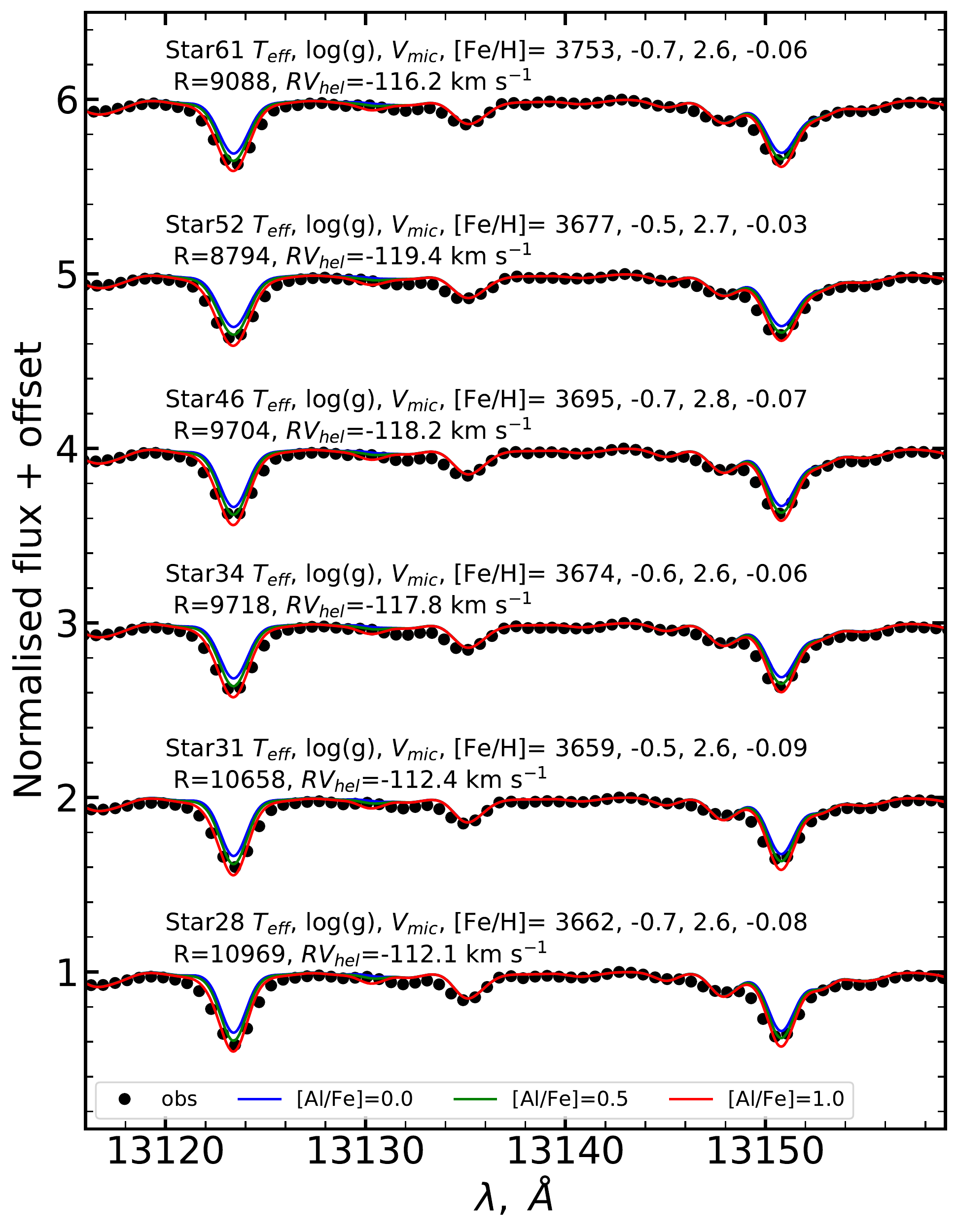}
\caption{Comparing Al spectral line for each RSG in our sample with models of [Al/Fe] = 0.0, 0.5 and 1.0 respectively while keeping all other parameters derived in the previous section constant.}
\label{Aluminum}
\end{figure*}

\begin{figure*}[ht!]
\plotone{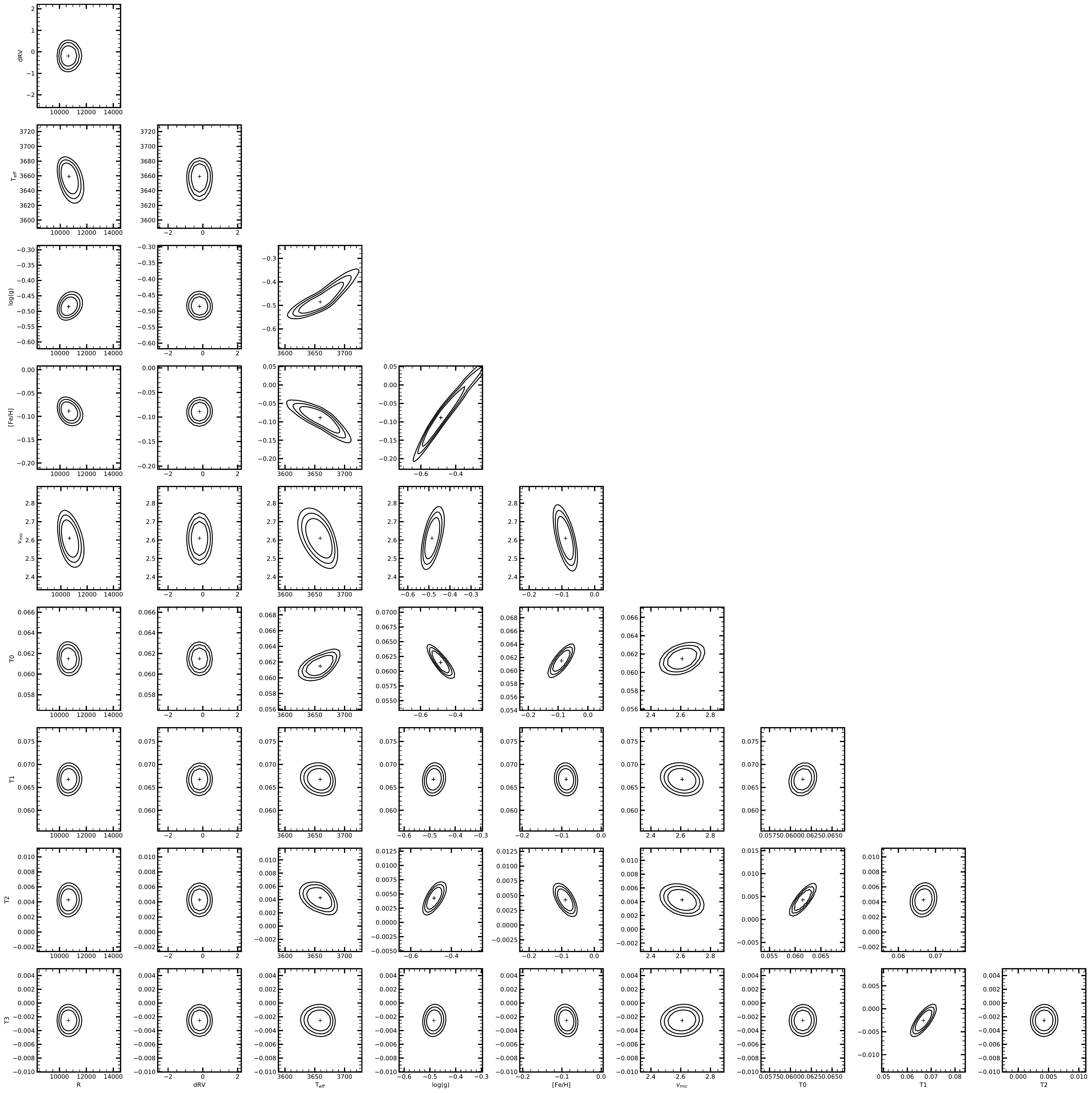}
\caption{The contour plot showing degeneracies between parameters. They were obtained by changing two parameters at a time and keeping the rest fixed to the optimal values. Contour lines are shown for values corresponding to 1, 2, 3 $\sigma$ levels. Optimal parameters are shown with pluses.}
\label{Fig2b}
\end{figure*}

It is well established now that globular clusters (GC) host multiple stellar populations
\citep[MSP; e.g.,][and references therein]{Milone13, Milone15, Milone16, Milone17a, Milone20, Bastian15, Bastian18, Bastian19, Bastian20} inferred through star-to-star variations in the abundances of some light elements (e.g., He, C, N, O, Na, Al). Several scenarios have been proposed to explain this phenomenon, with most implying multiple epochs of star-formation within the cluster, however none have fully succeeded to reproduce the increasing number of observations obtained in the past decade. Hence, the origin of this phenomenon is still debated. Even more puzzling is the fact that no evidence of MSPs has been found so far in lower-mass ($<$ 10$^4$ M$_\odot\/$) open clusters nor in 1-2 Gyr massive ($>$ 10$^5$ M$_\odot\/$) clusters \citep{Bastian18}.\\
\citet{Marco14} reports that VdBH 222 is almost certainly the most massive cluster observable in the U and B bands in the Milky Way. This makes it an ideal candidate to search for MSP, as several studies showed the there is a threshold of cluster mass for which MSPs can be observed \citep{Carretta10, Milone17a, Bastian19}.\\
In order to search for star-to-star light element abundance variations and quantitatively investigate the phenomenon of MSP in YMCs, we look for Al variations as Al lines in the wavelength range $13120-13155$\,\AA ${}$ have been previously used in the literature to look for the presence of MSPs in the spectra of RSGs in young clusters \citep{Cabrera-Ziri16, Lardo17}. \\
\citet{Pancino17} found that the extension of the Mg-Al anti-correlation (i.e. Al enhancement) depends on both metallicity and mass.
\citet{Cabrera-Ziri16} defined the [Al/Fe] spread as $\Delta$[Al$/$Fe]= mean([Al/Fe]) $-$ min([Al/Fe]) and divided the $\Delta$[Al$/$Fe] observed in GCs in three broad ranges: moderate, intermediate and extreme according to their $\Delta$[Al/Fe] values $\Delta$[Al/Fe]= 0.1, 0.3 and 0.7 dex, respectively.\\
We examine the Al variation in our cluster based on these ranges. The average best fitting values of [Al$/$Fe] for our RSG sample is 0.9 dex. In Figure \ref{Aluminum} we show our observations overplotted with models of [Al/Fe] = 0.0, 0.5 and 1.0 respectively. \\
$\Delta$[Al/Fe] for our sample is 0.1 which is in the moderate range defined by \citet{Cabrera-Ziri16}.
We infer that there is no evidence for MSP in this cluster, which is consistent with findings of other studies for clusters of this young age. As pointed by \citet{Cabrera-Ziri16}, this might be due to the fact that MPs only manifest themselves in low mass stars due to some evolutionary mechanism.

\section{Conclusion}

There are young massive clusters recently discovered in the central regions of the Milky Way that are still poorly characterised, requiring further study. One of them is our cluster VdBH 222.\\
Using (VLT)+X-shooter IR spectra ($11600-12200$\,\AA) of six RSGs in this clusters we apply the J-band full spectrum technique to derive the stellar parameters and abundances. Our results are summarized as follows:\\
1. The average radial velocity for our sample is $-$116.5$\pm$3 km s$^{-1}$ which is consistent with previous studies. \\
2. The RSGs of our sample have temperatures in the 3650$-$3750 K range. The order of the temperature values (hottest to coolest) are consistent with the spectral classes identified by \citet{Marco14} for this sample. \\
3. Our sample has surface gravity log(g) in the $-$0.7 to $-$0.5 range and microturbulence in the 2.6 to 2.8 km s$^{-1}$ range.\\
4. We provide the first [Fe$/$H] estimate for this YMC. We find the average [Fe$/$H] to be $-$0.07$\pm$0.02 which is in line with findings from other studies for clusters in the central region of the MW. \\
5. In our search for Multiple Stellar Populations in this YMC, we exclude at high confidence extreme [Al/Fe] enhancements similar to those observed in GCs, hence we infer that this massive, extremely young, open cluster does not show MSPs.
This is in line with other studies, as no evidence for MSPs has been yet observed in clusters younger than 2 Gyrs \citep[][and references therein]{Bastian18}. However,
we need more observations in order to better understand if this is a property attributable to all young massive clusters or it is because MSPs only manifest themselves in low mass stars due to some evolutionary mechanism. The origin of MSPs is still unclear and further studies of young massive clusters may provide constraints for better understanding this phenomenon.\\
Our analysis of this cluster is based on the mass and age derived by \citet{Marco14}, where the cluster's total mass was determined from its similitude with objects having comparable number of RSGs, not from star counts or other methods for membership determinations (like proper motions, parallaxes ... etc). VdBH 222 was chosen for this study although it has not been previously studies comprehensively, because it is one of the few clusters with confirmed multiple RSGs. The uncertainty in the total cluster mass, age and distance should be noted. More accurate studies on this cluster are needed.

\section*{Acknowledgements}

We thank the referee for the careful review, which helped improve the manuscript.
During part of this work, R. Asa'd was a visiting scientist at ESO-Garching whose hospitality is greatly acknowledged.
This work is based on observations collected with VLT/X-shooter under ESO programme 0103.D$-$0881(A) (PI R. Asad).
This work is supported by the EFRG18-SET-CAS-74, FRG19-S-S131 and OAP-CAS-054 grants P.I., R. Asa’d from American University of Sharjah. This paper represents the opinions of the authors and does not mean to represent the position or opinions of the American University of Sharjah.

\bibliographystyle{aasjournal}
\bibliography{references}

\clearpage
\begin{table*}
\caption{Parameters and Abundances of the RSGs sample}
\label{Targets_V222}
\begin{tabular}{ccccccccccc}
\hline
ID & Res. & RV (km s$^{-1}$) & RA & Dec & T$_{eff} (K)$ & log(g) &  [Fe$/$H] & $\xi$ & S$/$N \\
\hline
Star28 &10969&-112.1&259.70310&-38.29149& 3662$\pm$21& $-$0.7$\pm$0.1& $-$0.08$\pm$0.04& 2.6$\pm$0.1 & 291  \\
Star31 &10658&-112.4&259.70198&-38.28356& 3659$\pm$27& $-$0.5$\pm$0.1& $-$0.09$\pm$0.06& 2.6$\pm$0.1 & 261 \\
Star34 &9718&-117.8&259.68680&-38.29786& 3674$\pm$21& $-$0.6$\pm$0.1& $-$0.06$\pm$0.04& 2.6$\pm$0.1 & 118 \\
Star46 &9704&-118.2&259.69237&-38.27003& 3695$\pm$21& $-$0.7$\pm$0.1& $-$0.07$\pm$0.05& 2.8$\pm$0.1 & 244 \\
Star52 &8794&-119.4&259.70270&-38.30095& 3677$\pm$21& $-$0.5$\pm$0.1& $-$0.03$\pm$0.04& 2.7$\pm$0.1 & 251 \\
Star61 &9088&-116.2&259.69552&-38.28856& 3753$\pm$22& $-$0.7$\pm$0.1& $-$0.06$\pm$0.04& 2.6$\pm$0.1 & 234 \\
\hline
Mean Value & - & $-$116.5$\pm$3 & - & - & 3686$\pm$35 & $-$0.6$\pm$0.1 & $-$0.07$\pm$0.02 & 2.7$\pm$0.1 & 233 \\
\hline
\end{tabular}
\end{table*}
\label{T1}

\end{document}